\documentclass[twocolumn,showpacs,prl]{revtex4}

\usepackage{amssymb}
\usepackage{graphicx}
\usepackage{dcolumn}
\usepackage{bm}

\begin{document}


\title{Laser field absorption in self-generated electron-positron pair plasma}
\author{E.~N.~Nerush, I.~Yu.~Kostyukov}
\email{kost@appl.sci-nnov.ru}
\affiliation{Institute of Applied Physics, Russian Academy of Sciences, 603950 Nizhny Novgorod, Russia}
\author{A.~M.~Fedotov, N.~B.~Narozhny,}
\affiliation{National Research Nuclear University MEPhI, Moscow, 115409, Russia}
\author{N.~V.~Elkina, H.~Ruhl}
\affiliation{Ludwig-Maximillians Universit\"{a}t M\"unchen, 80539, Germany}


\begin{abstract}
Recently much attention has being attracted to the problem of limitations
on the attainable intensity of high power lasers [A.M.~Fedotov {\it
et al.} Phys.~Rev.~Lett. \textbf{105}, 080402 (2010)]. The laser
energy can be absorbed by electron-positron pair plasma produced from 
a seed by strong laser field via development of the electromagnetic
cascades. The numerical model for self-consistent study
of electron-positron pair plasma dynamics is developed. Strong
absorption of the laser energy in self-generated overdense
electron-positron pair plasma is demonstrated. It is shown that the
absorption becomes important for not extremely high laser intensity
$I \sim 10^{24}$~W/cm$^2$ achievable in the nearest future.
\end{abstract}

\maketitle

Due to an impressive progress in laser technology, laser pulses with
peak intensity of nearly $2 \times 10^{22}$ W/cm$^2$ are now
available in the laboratory \cite{Michigan2008}. When the matter is
irradiated by so intense laser pulses ultrarelativistic dense plasma
can be produced. Besides of fundamental interest, such plasma is an
efficient source of  particles and radiation with extreme parameters
that opens bright perspectives in development of advanced particle
accelerators \cite{Esarey2009}, next generation of radiation sources
\cite{Kiselev2004,Rousse2004}, laboratory modeling of astrophysics
phenomena \cite{Remington2006}, etc. Even higher laser intensities
can be achieved with the coming large laser facilities like ELI
(Extreme Light Infrastructure) \cite{eli} or HiPER (High Power laser
Energy Research facility) \cite{hiper}. At such intensity the
radiation reaction and quantum electrodynamics (QED) effects become
important \cite{Tajima2002,Mourou2006,Nerush2007,Bell2008,Fedotov2010,Sokolov2010}.

One of the QED effects, which has recently attracted much
attention, is the electron-positron pair plasma (EPPP) creation in a
strong laser field \cite{Bell2008,Fedotov2010}. The plasma can be
produced via avalanche-like electromagnetic cascades: the seed
charged particles are accelerated in the laser field, then they emit
energetic photons, the photons by turn decay in the laser field and
create electron-positron pairs. The arising electrons and positrons
are accelerated in the laser field and produce new generation of the
photons and pairs. It is predicted \cite{Fedotov2010} that
an essential part of the laser energy is spent on EPPP production and heating. 
This can limit the attainable intensity of high power lasers. That prediction
was derived using simple estimates, therefore self-consistent
treatment based on the first principles is needed.

The collective dynamics of EPPP in strong laser field is a very
complex phenomenon and numerical modeling becomes important to
explore EPPP. Up to now the numerical models for collective QED
effects in strong laser field have been not self-consistent. One
approach in numerical modeling is focused on plasma dynamics and
neglects the QED processes like pair production in the laser field.
It is typically based on particle-in-cell (PIC) methods and uses
equation for particle motion with radiation reaction forces taken
into account \cite{Sokolov2010}. The second one is based on Monte
Carlo (MC) algorithm for photon emission and electron-positron pair
production. This approach has been used to study the dynamics of
electromagnetic cascades \cite{Anguelov1999}. However, it completely
ignores the self-generated fields of EPPP and the reverse effect of
EPPP on the external field. The latter effect is especially important
to determine the limitations on the intensity of high power lasers
\cite{Fedotov2010,Bulanov2010}.

Quantum effects in strong electromagnetic fields can be
characterized by the dimensionless invariants
\cite{Nikishov1964,Landau4} $\chi_e = e\hbar / (m^3c^4) |F_{\mu \nu
} p_\nu | \approx \gamma ( F_ \bot / eE_{cr} )$  and $\chi _\gamma
\approx (\hbar \omega / mc^2 ) (F_ \bot / eE_{cr})$, where $F_{\mu
\nu } $ is the field-strength tensor, $p_\mu $ is the particle
four-momentum, $\hbar \omega $ is the photon energy, $\gamma $ is
the electron gamma-factor, $F_ \bot $ is the component of Lorentz
force, which is perpendicular to the electron velocity, $E_{cr} =
m^2c^3 / (e\hbar ) = 10^{16}$~$\mbox{V/cm}$ is the so-called QED
characteristic field, $\hbar$ is the Planck constant. $\chi_e$
determines photon emission by relativistic electron while $\chi
_\gamma $ determines interaction of hard photons with
electromagnetic field. QED effects are important when $\chi_e
\gtrsim 1$ or $\chi _\gamma \gtrsim 1$. If $\chi_e \gtrsim 1$ then
$\hbar \omega \sim \gamma m c^2 $ and the quantum recoil imposed on the
electron by the emitted photon is strong. The probability rate of
emission of a photon with energy $\hbar \omega$ by relativistic
electron with gamma-factor $\gamma $ can be written in the form
\cite{Baier1998,Landau4,Elkina2010}
\begin{eqnarray}
 dW_{em} \left(  \xi \right)  & = & \frac{\alpha m c^2}{\sqrt{3}
 \pi  \hbar \gamma } \left[ \left( 1 - \xi + \frac{1}{1 - \xi }
 \right) K_{2 / 3} ( \delta ) \right. \nonumber \\
&& \left. - \int_\delta ^\infty {K_{1 / 3} \left( s \right)} ds
\right] d \xi ,
\label{dwem}
\end{eqnarray}
\noindent where $\hbar \omega $ is the photon energy, $m$ is the
electron mass, $c$ is the speed of light, $\delta = 2\xi /[ 3(1 - \xi
)\chi_e ]$ and $\xi = \hbar \omega / (\gamma mc^2)$. $\hbar \omega
dW_{em} $ can be considered as the energy distribution of the
electron radiation power. For electron radiation in constant
magnetic field ${\bf B} $ perpendicular to the electron velocity it
reduces to the synchrotron radiation spectrum in the classical limit
$\chi_e \ll 1$ \cite{Landau2,Baier1998}. The probability rate of
electron-positron pair production by decay of a photon with energy
$\hbar \omega$  is \cite{Landau4,Baier1998,Elkina2010}
\begin{eqnarray}
 dW_{pair} \left(  \eta _- \right)  & = & \frac{\alpha m^2 c^4}
 {\sqrt{3} \pi  \hbar ^2 \omega }\left[ \left( \frac{\eta _+}{\eta _-}
 + \frac{\eta _-}{\eta _+} \right) K_{2 / 3} ( \delta ) \right.
 \nonumber \\
&& \left. - \int_\delta ^\infty {K_{1 / 3} \left( s \right)} ds \right] d \eta ,
\label{dpair}
\end{eqnarray}
\noindent where $\delta = 2 / (3 \chi _\gamma  \eta _-  \eta _+ )$,
$\eta_- = \gamma m c^2 / (\hbar \omega)$ and $\eta _+ = 1 - \eta _-$
are the normalized electron and positron energies, respectively. It
follows from Eq.~(\ref{dpair}) that in the classical limit $\chi
_\gamma \ll 1$ this probability is exponentially small.

To study EPPP dynamics we have developed two-dimensional numerical
model based on PIC/MC methods. Similar methods has been used
previously for modeling of discharges in gases \cite{Shklyaev2009}.
Recently a one-dimensional PIC/MC model has been developed to
simulate pair cascades in magnetosphere of neutron stars
\cite{Timokhin2010}. However the latter model is electrostatic and
the classical approach is used for photon emission with radiation
reaction force in the equation of motion. In our numerical model we
use more general approach. We exploit the fact that there is a large
difference between the photon energy scales in EPPP. The photon energy of the
laser and plasma fields is low ($\hbar \omega \ll m c^2$) while the
energy of the photons emitted by accelerated electrons and positrons
is very high ($\hbar \omega \gg m c^2$). The emitted photons are
hard and can be treated as particles. Conversely, the evolution of the 
laser and plasma fields is calculated by numerical solution of the
Maxwell equations. Therefore, the dynamics of electrons, positrons
and hard photons as well as the evolution of the plasma and laser
fields are calculated by PIC technique while emission of hard
photons and pair production are calculated by MC method.

The photon emission is modeled as follows. On every time step for
each electron and positron we sample a photon emission by a
probability distribution which approximates Eq.~(\ref{dwem}) with
the accuracy within $5\%$. The new photon is included in the
simulation region. The coordinates of a new photon are equal to the
electron (positron) coordinates at the emission instance. The photon
momentum is parallel to the electron (positron) momentum. The
electron (positron) momentum value is decreased by the value of the
photon momentum. Similar algorithm is used for modeling of
pair production by photons. The new electron and positron are
added in the simulation region while the photon that produced a
pair is removed. The sum of the electron and positron energy is
equal to the photon energy. The pair velocity is directed
along the photon velocity at the instance of creation.

The MC part of our numerical model has been benchmarked to
simulations performed by other MC codes. We simulated the
electromagnetic showers in a static homogeneous magnetic field, the
interaction of relativistic electron beam with a strong laser pulse,
and the development of electromagnetic cascades in circularly polarized
laser pulses. The obtained results are in reasonably good agreement
with those published by other authors and are discussed in
Ref.~\cite{Elkina2010}. The particle motion and evolution of the
electromagnetic field are calculated with standard PIC technique
\cite{Pukhov1999}. The PIC part of the model
is two-dimensional version of the model used in Ref.~\cite{Nerush2009}.
In order to prevent memory overflow during
simulation because of the exponential growth of particle number in a
cascade, the method of particles merging is used
\cite{Timokhin2010}. If the number of the particles
becomes too large the randomly selected particles are deleted while the charge, 
mass, and the energy of the rest particles are increased by the
charge, mass, and energy of the deleted particles, respectively.

We use our numerical model to study production and dynamics of EPPP
in the field of two colliding linearly polarized laser pulses. The
laser pulses have Gaussian envelopes and propagate along the
$x$-axis. The components of the laser field at $t=0$ are
$E_y,B_z = a_0 \exp ( -y^2/\sigma_r^2 ) \sin \zeta \left[
e^{-(x+x_0)^2/\sigma_x^2 } \pm e^{- ( x-x_0)^2/\sigma_x^2 }
\right]$, where the field strengths are normalized to $mc\omega _L/
|e|$, the coordinates are normalized to $c/\omega_L$, time is
normalized to $1/\omega _L$, $a_0 = |e| E_0 / (mc\omega _L)$, $E_0$
is the electric field amplitude of a single laser pulse, $\omega_L$
is the laser pulse cyclic frequency, $2 x_0$ is the initial distance
between the laser pulses, $\zeta = x - \phi$, and $\phi$  is the
phase shift. The parameters of our simulations are $\phi=0.8 \pi$,
$a_0 = 1.2\cdot 10^3$, $\sigma_x =125$, $\sigma_r = 40$, $x_0 =
\sigma_x/2$ that for the
wavelength $\lambda = 2 \pi c / \omega _L = 0.8 \text{ } \mu
\text{m}$ correspond to the intensity  $3 \cdot 10^{24}
\text{ W/cm}^2$, pulse duration  $100 \text{ fs}$, the focal spot
size $10 \text{ } \mu \text{m}$ at $1/e^2$ intensity level.
The cascade is initiated by a single electron
 located at $x=y=0$ with zero initial momentum for $t=0$ when
the laser pulses approach to each other (the distance between
the pulse centers is $\sigma_x$).

\begin{figure}
\includegraphics[width=8cm,clip]{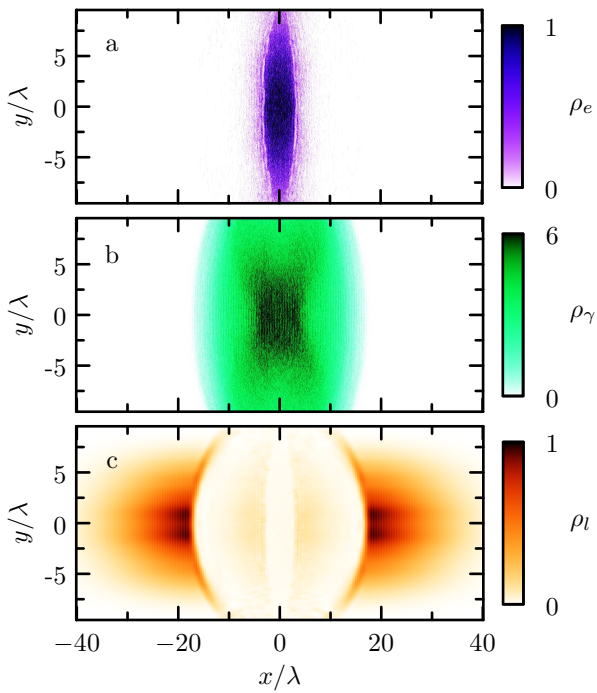}
\caption{The normalized electron density $\rho _e = n_e / (a_0 n_ {cr})$ (a), 
the normalized photon density $\rho _\gamma = n_\gamma / (a_0 n_ {cr})$ (b)
 and the laser intensity normalized to the maximum of the initial intensity $\rho _l$ (c) 
 during the collision of two linearly-polarized laser pulses at $t=25.5 \lambda/c$.}
\label{movie1}
\end{figure}

\begin{figure}
\includegraphics[width=8cm,clip]{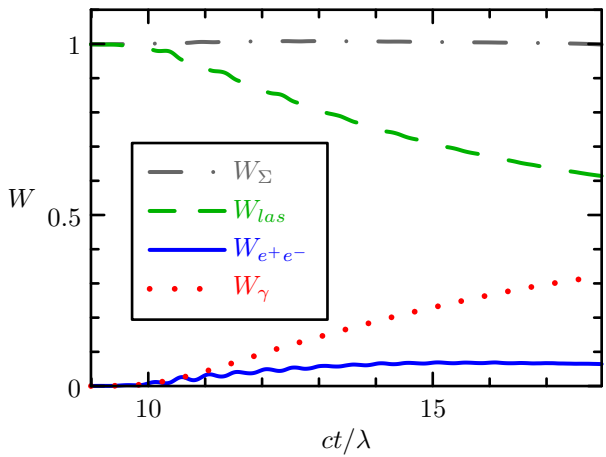}
\caption{The electron and positron energy (solid line), the
photon energy (dotted line), the laser energy (dashed line) and the
total energy of the system (dash-dotted line) as functions of time.
All the energies are normalized to the initial energy of the system.}
\label{energy}
\end{figure}

The later stage ($t=25.5 \lambda/c$) of the cascade development is
shown in Fig.~\ref{movie1}, where the electron and photon
density distributions and the laser intensity distribution are
presented. The laser pulses passed through each other by this
time instance and the distance between the pulse centers becomes about
$1.6 \sigma_x$. It is seen from Figs.~\ref{movie1} that the micron-size
cluster of overdense EPPP is produced and the laser energy at the backs
of the incident laser pulses is spent on EPPP production and heating.
The plasma density exceeds the relativistic critical density
$a_0 n_{cr}$ in about $2$ times, where $n_{cr} = m\omega^2/ \left( 8 \pi e^2 \right)$ 
is the nonrelativistic critical density for the electron-positron plasma.
The evolution of the particle and laser energy is shown in
Fig.~\ref{energy}. It is seen from Fig.~\ref{energy} that about
a half of the laser energy is absorbed by self-generated EPPP
and then mostly reradiated in ultrashort pulse of gamma-quanta. The
total energy of the particles in the cascade and the electromagnetic
field is conserved with accuracy about $1\%$ during our simulation.

\begin{figure}
\includegraphics[width=8cm,clip]{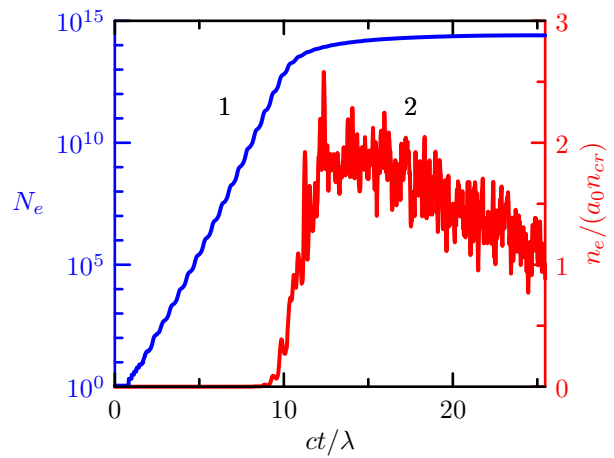}
\caption{The number of the electrons produced in the cascade (line 1)
and the EPPP density normalized to the relativistic critical density
(line 2) as functions of time.}
\label{nvst}
\end{figure}

\begin{figure}
\includegraphics[width=8cm,clip]{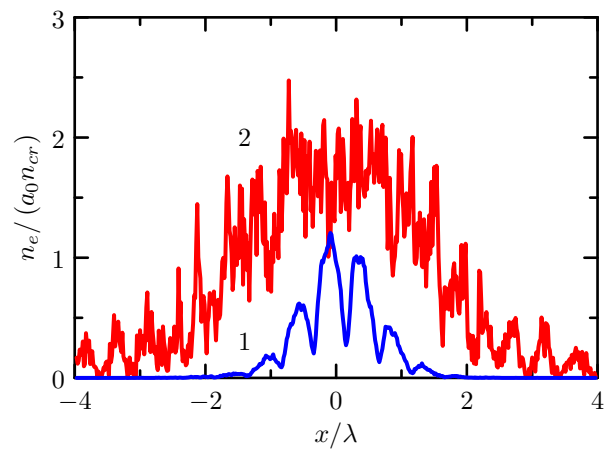}
\caption{The profile of the electron density along $x$-axis at $y=0$
for initial stage $c t=6.4 \lambda $ (line 1) and for the later stage
$c t=16.6 \lambda$ (line 2) of the cascade development. The electron
density for $c t=16.6 \lambda $  is normalized to  $a_0 n_{cr}$
and that for $c t=6.4 \lambda $ is normalized to  $3.3 \times 10^ {-6} a_0 n_{cr}$.}
\label{nvsx}
\end{figure}

\begin{figure}
\includegraphics[width=8cm,clip]{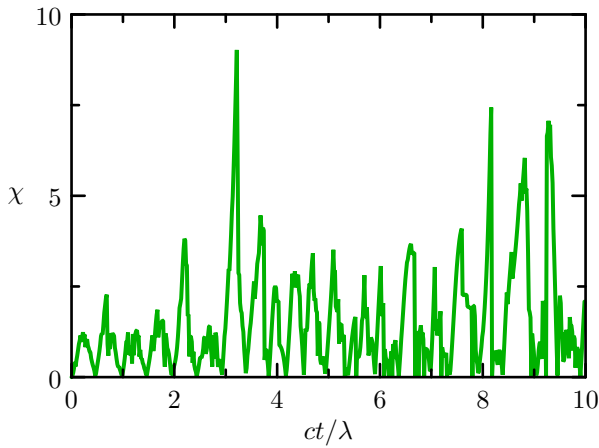}
\caption{The dependence of $\chi _e$ for the primary electron on time.}
\label{chi}
\end{figure}

At initial stage of the cascade development the number of created
particles is growing exponentially $N \sim e^{\Gamma t}$ \cite{Fedotov2010}, 
where $\Gamma$ is the multiplication rate. It follows from the energy
conservation law that the number of particles that can be created is
limited by the laser pulses energy. Thus, at some instant the
exponential growth is replaced by much slower growth.
Equating the initial energy of laser pulses
to the overall particles energy after the pulse collision we get $ N
\sim a_0^2 \sigma_x \sigma_r^2 N_0 / \bar \gamma$, where
we assume $N_e \sim N_p \sim N_{ph} \sim N$, $N_e$, $N_p$ and $N_{ph}$ are
the number of electrons, positrons and photons produced by the
cascade, respectively, $mc^2 \bar
\gamma$ is the average particle energy, $N_0 = n_{cr}(c/\omega)^3 =
\lambda/( 16\pi r_e )$, $r_e = e^2/(mc^2) $. The multiplication rate
decreases when the field strength goes down, that, by turn, occurs
if the plasma density reaches the value $a_0 n_{cr}$. This is in
good agreement with the numerical results shown in Fig.~\ref{nvst},
where the multiplication rate drops dramatically and EPPP density
reaches the value about $a_0 n_{cr}$ at the same instant of time
$t_s \approx 10 \lambda/c$. The value of $t_s$ can be estimated as
$t_s \approx \Gamma^{-1} \ln N$.  It follows from Fig.~\ref{nvst}
that $\Gamma \approx 0.6 \omega _L$ for $t<t_s$. The typical
lifetime $t_{em}$ for electrons and positrons with respect to hard
photon emission can be estimated as $1/\Gamma>1/\omega _L$
\cite{Fedotov2010}. Thus, for the parameters of numerical simulation
$\bar \gamma$ can be estimated as $\bar \gamma \sim a_0$ hence $N
\sim a_0 \sigma_x \sigma_r^2 N_0 \sim 4\cdot 10^{14}$ and $c t_s /
\lambda \sim 9$ that are in good agreement with the corresponding
values from Fig.~\ref{nvst}.

It is shown in Ref.~\cite{Bulanov2010} that the cascades do not
arise in $B$-node of linearly polarized standing electromagnetic
wave so far as $\chi_e$ and $\chi _\gamma$ are less than unity.
However, our numerical simulations show that the cascade quasi-periodically 
develops between $B$ and $E$ nodes of such a wave. This is because under such
conditions the electron motion becomes complicated and is not
confined to the direction of polarization on the temporal scales
about the laser period. It turns out that there occur the time
intervals of duration of the order of $\omega _L^{-1}$ with $\chi _e
> 1$ (see Fig.~\ref{chi}) on which the cascade can develop. 
The time modulations of $N_e (t)$ and of EPPP density along the $x$-axis at
the initial stage of the cascade development are seen in
Fig.~\ref{nvst} and \ref{nvsx}, respectively. At the later stages
the spatial  modulation of the density is strongly smoothed out due to EPPP
expansion (see  Fig.~\ref{nvsx}, line 2).

In conclusion we develop the numerical model which allows us to
study EPPP dynamics in strong laser field self-consistently. We have
demonstrated efficient production of EPPP at
the cost of the energy of the laser pulses. We show that even not
extremely high intensity laser pulses ($I\sim 10^{24}$~W/cm$^2$ with
duration $\sim 100$~fs) can produce overdense EPPP so that the QED effects
can be experimentally studied with near
coming laser facilities like ELI \cite{eli} and HiPER \cite{hiper}.
The simulations and estimates show that for intensity
$I>10^{26}$~W/cm$^2$ the overdense EPPP can be produced during a
single laser period. In such high-intensity regime few-cycle laser
pulses can be used in experiments. High-energy photons or electron-positron
pair can be also used as a seed to initiate cascade instead of an electron.
Photon-initiated cascade can be more suitable for experimental
study in low intensity regime ($I\sim 10^{24}$~W/cm$^2$) because the
laser intensity threshold for pair creation in vacuum is about $\sim
10^{26}$~W/cm$^2$ \cite{Narozhny}.

\begin{acknowledgments}
This work was supported in parts by the Russian Foundation for Basic
Research, the Ministry of Science and Education of the Russian
Federation, the Russian Federal Program ``Scientific and
scientific-pedagogical personnel of innovative Russia'', the grant
DFG RU 633/1-1, and the Cluster-of-Excellence 'Munich-Centre for
Advanced Photonics' (MAP).
\end{acknowledgments}

\end{document}